\documentclass[twocolumn,prl,aps,twocolumn]{revtex4}
\usepackage[latin9]{inputenc}
\usepackage{color}
\usepackage{amstext}
\usepackage{amssymb}
\usepackage{graphicx}
\usepackage[unicode=true,pdfusetitle,
 bookmarks=false,
 breaklinks=false,pdfborder={0 0 1},backref=false,colorlinks=false]
 {hyperref}
\hypersetup{
 bookmarksnumbered=false,bookmarksopen=false}

\usepackage{amsmath}
\usepackage{graphicx}
\usepackage{amssymb}
\usepackage{txfonts}

\begin{document}

\title{Pulse-phase control for spectral disambiguation in quantum sensing protocols}

\author{J. F. Haase}
\affiliation{Institut f\"ur Theoretische Physik and IQST, Albert-Einstein-Allee 11, Universit\"at
Ulm, D-89069 Ulm, Germany}
\author{Z.-Y. Wang}
\affiliation{Institut f\"ur Theoretische Physik and IQST, Albert-Einstein-Allee 11, Universit\"at
Ulm, D-89069 Ulm, Germany}
\author{J. Casanova}
\affiliation{Institut f\"ur Theoretische Physik and IQST, Albert-Einstein-Allee 11, Universit\"at
Ulm, D-89069 Ulm, Germany}
\author{M. B. Plenio}
\affiliation{Institut f\"ur Theoretische Physik and IQST, Albert-Einstein-Allee 11, Universit\"at
Ulm, D-89069 Ulm, Germany}

\begin{abstract}
We present a method to identify spurious signals generated by finite-width pulses in quantum sensing experiments and apply it to recently proposed dynamical decoupling sequences for accurate spectral interpretation. We first study  the origin of these fake resonances and quantify their behavior in a situation that involves the measurement of a classical magnetic field. Here we show that a change of the initial phase of the sensor or, equivalently, of the decoupling pulses leads to oscillations in the spurious signal intensity while the real resonances remain intact. Finally we extend our results to the quantum regime for the unambiguous detection of remote nuclear spins by utilization of a nitrogen vacancy sensor in diamond.
\end{abstract}
\maketitle

\section{Introduction}
In current quantum sensing experiments involving nitrogen vacancy (NV) centers in diamond~\cite{Doherty13, Dobrovitski13, Wu2016}, dynamical decoupling (DD) pulse sequences such as Carr-Purcell-Meiboom-Gill (CPMG)~\cite{Carr54,Meiboom58}, or the \textit{XY} family~\cite{Maudsley86,Gullion90, Ahmed13}  are used to design filter functions~\cite{Kofman2001,Biercuk2009,Sousa2009} only transmissive for particular frequencies by refocusing the undesired couplings. The operating principle to detect an external signal, either classical or quantum~\cite{Balasubramanian2008},  corresponds to having the NV center, i.e., the quantum sensor, evolving under the action of these decoupling pulses and the signal. Whenever the generated filter is permeable for a certain frequency component of the signal, the quantum sensor gathers a phase that will be subsequently  measured leading to a spectrum that characterizes its environment~\cite{Kolkowitz12, London13, Muller14, Cywinski2008, Zhao2012}.
 
A filter function is created by a sequence of microwave $\pi$-pulses applied on the NV center. 
For standard DD sequences such as the CPMG or the \textit{XY} family, the expected resonances can only occur at the frequencies $l\omega_{\text{DD}}$, where $l$ are odd integers and $\omega_{\text{DD}}=\pi/t_{\text{free}}$ for a pulse interval of $t_{\text{free}}$~\cite{Kolkowitz12, Zhao2012}. In the same manner DD schemes employing composite pulses admit a similar description~\cite{Souza12}. However it has been recently shown~\cite{Loretz15} that, due to the finite width of the applied pulses, the quantum sensor still accumulates a phase if $l \omega_{\text{DD}}/k$ matches the signal frequency $\omega_{\text{ac}}$ or, equivalently,  $\omega_{\text{DD}} = \alpha \omega_{\text{ac}}$ with $\alpha=k/l$. Here $k \in  \mathbb{N}$ with the maximum value of $k$  defined by the outer period of the sequence, and odd numbers of $k$  are excluded by symmetric sequences~\cite{Loretz15}. Therefore the spurious responses with $k\neq 1$ lead to spectral ambiguities and to a misinterpretation of the signals present in the environment. In particular, the $k=4$ spurious resonance of a $^{13}$C spin may be falsely interpreted as the $k=1$ resonance of a hydrogen spin.  

In this work we show that spurious responses in the measured spectrum can be identified and separated from the real ones by controlling  the initial phase of the quantum sensor or the phase of the decoupling pulses. More specifically, we show how the intensity of the spurious peaks changes when we vary this phase  while the real peaks do not change in the spectrum. Furthermore we show how this method can be combined with recently proposed robust DD sequences for an accurate characterization of the spin environment.  

The article is organized as follows: In Sec.~\ref{motivation} we motivate our method by studying the spurious signals' behavior in the detection of classical fields.  
In Sec.~\ref{quantum} we apply the method to the quantum regime where, in particular, we will make use of an NV center in diamond as a  quantum sensor. Furthermore we will combine our protocol with the recently proposed adaptive XY (AXY) DD pulse sequences for accurate spin detection~\cite{Casanova15, Wang15}.

\section{Theory}\label{motivation}
\subsection{Detection of a classical signal}

To understand the presence and detection of spurious resonances we consider a sensor spin subjected to a static magnetic field, $\vec{B} = B_z \ \hat{z}$, and driven by a classical ac-field, i.e., the external signal, applied in the same $\hat{z}$ direction with angular frequency $\omega_{\text{ac}}$ and amplitude $B$. For the case of an NV based sensor, we choose the $\hat{z}$-direction along the NV symmetry axis. In addition we consider the action of microwave $\pi$-pulses  for both coherence protection of the sensor spin and detection of the ac-field. The relevant Hamiltonian in a rotating frame with respect to the static $B_z$ field reads ($\hbar=1$)
\begin{eqnarray}\label{classical}
H(t) &=& \gamma_n B \sin (\omega_{\text{ac}} t +\theta)\frac{\sigma_z}{2} - \Delta \frac{\sigma_z}{2} + H_c,
\label{eq:Hamiltonian}
\end{eqnarray}
where $\sigma_\mu,\mu=x,y,z$ are Pauli matrices, $\theta$ is the initial phase of the ac-field, and $\Delta$ a possible detuning of the driving field. The control Hamiltonian 
\begin{eqnarray}
H_c = \frac{1}{2}\Omega [ \cos{(\varphi_i+\vartheta)} \ \sigma_x + \sin{(\varphi_i+\vartheta)} \sigma_y ] 
\label{eq:ControlHamiltonian}
\end{eqnarray}
is applied stroboscopically leading to the action of the decoupling $\pi$-pulses on the sensor spin. The pulse-phase $\varphi_i$ controls the rotation axis on the $x$-$y$ plane, while $\vartheta$ sets an overall phase on the pulses which we set to zero for the following calculations.

In the sensing protocol, the sensor spin is initialized in the state described by the density matrix 
\begin{eqnarray}
	\rho_0 = \frac{1}{2}\left(\begin{array}{cc} 1 & e^{-i\phi} \\ e^{i\phi} & 1\end{array} \right),
\end{eqnarray}
where $\phi$ corresponds to the initial phase of the state. After applying a DD pulse sequence, the density matrix of the central spin becomes $\rho(t)$ and we consider the transition probability $P = 1-\mathrm{Tr}[\rho(t)\rho_0]$ as the measured spectrum.

The effects of the control pulses and the ac-field on $\rho(t)$ can be described by a sequence of rotations 
\begin{eqnarray}
R_{\hat{n}}(\kappa) = e^{-i\kappa\hat{n}\cdot\vec{\sigma}/2}
\end{eqnarray}
on the central spin, where $\vec{\sigma}=(\sigma_x,\sigma_y,\sigma_z)^T$. 
Note that for instantaneous $\pi$ pulses $\kappa=\pi$. The effect of each instantaneous $\pi$ pulse around an axis lying in the $x$-$y$ plane corresponds to the change $\sigma_z \mapsto -\sigma_z$.  The free evolution between $\pi$-pulses gives rise to a phase accumulation
\begin{eqnarray}
\kappa_{\text{free},j}&=&\int_{t_j}^{t_{j+1}}\mathrm{d}\tau \gamma_n B\sin(\omega_{\text{ac}}\tau+\theta) \notag \\
&=& \frac{\gamma_n B}{\omega_{\text{ac}}}\left[\cos(\omega_{\text{ac}}t_j+\theta)-\cos(\omega_{\text{ac}}t_{j+1}+\theta)\right], \label{eq:ACSignal}
\end{eqnarray}
induced by the ac-field during the free evolution between the times $t_j$ and $t_{j+1}$ where pulses are applied. In this manner one can find that, for ideal control, the measured signal is  $P=\sin^2\left[\sum_{j=0}^N(-1)^j\kappa_{\text{free},j}/2\right]$ \cite{Chaudhry14}. This signal depends on the initial signal phase $\theta$. The effect of the detuning $\Delta$, which can be treated as static noise as shown in Eq.~(\ref{classical}), would be ideally removed by the DD sequence. For experiments where no control over $\theta$ is available, the signal would have to be averaged leading to a loss in contrast~\cite{Lange11}. \\

\subsection{Identifying spurious responses}
\begin{figure}[t!]
\begin{centering}
\vspace{0cm}
\hspace{-0.3cm} \includegraphics[width=0.95\columnwidth]{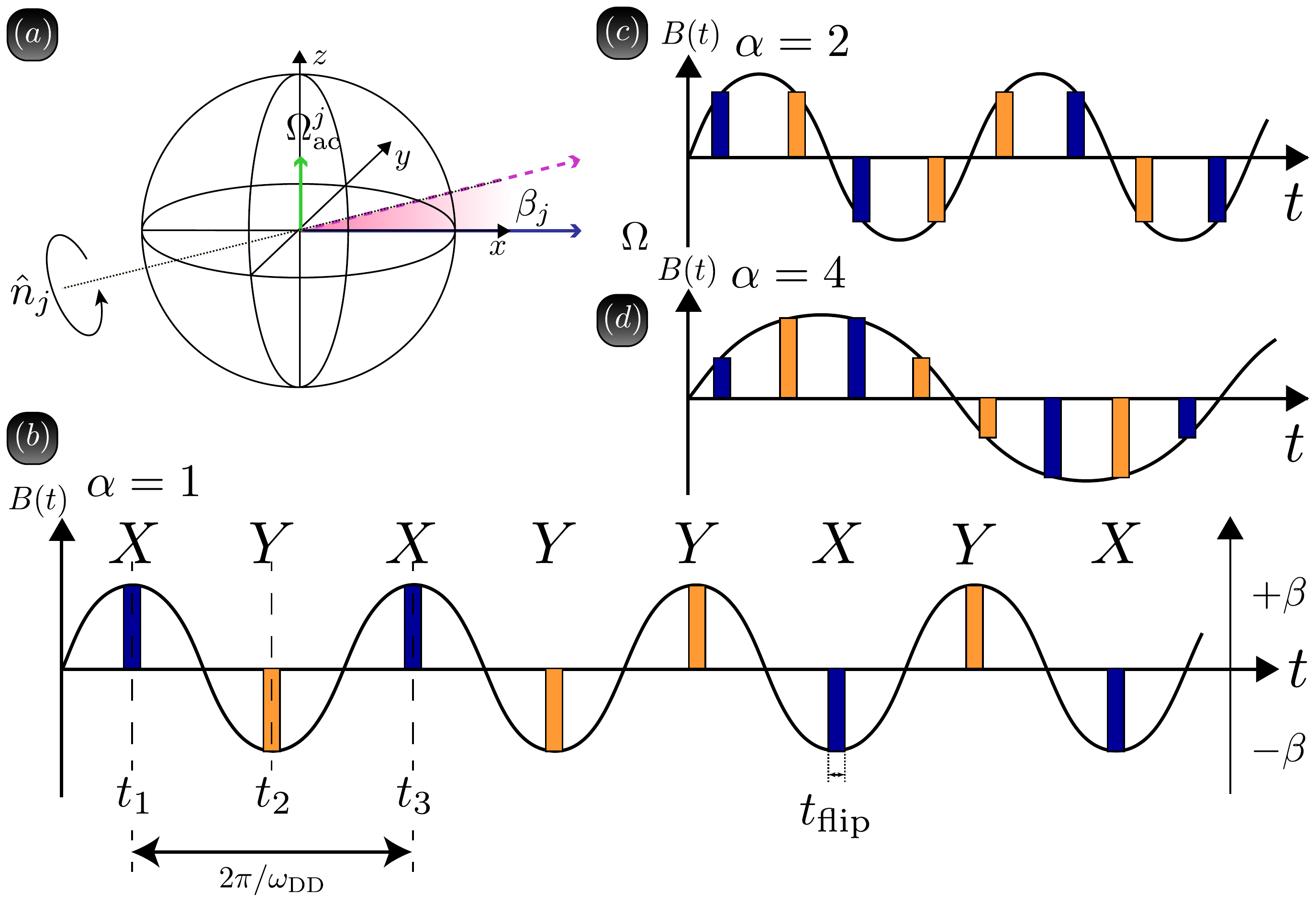}
\par\end{centering}
\caption{(Color online) (a) Visualization of the actual control axis at the present of signal fields on the Bloch sphere. The driving field $\Omega$ in the $x$-$y$ plane and the ac-field $\Omega_{ac}^{j}$ parallel to the $z$ axis add to a total driving field along $\hat{n}_{j}$, which set the angle $\beta_{j}$ out of the $x$-$y$ plane. (b) Locations of the X [blue (dark grey)] and Y [orange (light grey)] pulses with respect to the ac field for $\alpha=1$ and $\theta=0$. The height of each square pulse is proportional to the tilting angle $\beta$. (c) and (d) are the illustrations similar to (b) but for $\alpha=2$ and $\alpha=4$, respectively.}
\label{fig:harmonicaxis}
\end{figure}
The spurious resonances are caused by non-instantaneous $\pi$-pulses. To capture the physics of spurious resonances, we consider $\pi$-pulses with constant amplitudes and with a pulse duration $t_{\text{flip}} =\pi/\Omega$.
For the sake of simplicity on the following discussion we will assume $\Delta = 0$, see Eq.~\eqref{eq:Hamiltonian}. The presence of the ac-field  during $t_{\text{flip}}$  of the $j$-th pulse changes the rotation axis by an angle $\beta_j$ out of the $x-y$ plane, see Fig \ref{fig:harmonicaxis}(a). The value of $\beta_j$ is set by the relative strengths of the ac-field and the $j$-th pulse at time $t_j$, i.e., $\beta_j=\tan^{-1}\left[\gamma_n B\sin\left(\omega_{\text{ac}} t_j+\theta \right)/\Omega\right]$ (assuming that the ac field experiences almost no change during $t_{\text{flip}}$). In the following we consider a typical experimental situation where the signal amplitude, $\gamma_n B$, is small compared with $\Omega$ leading to $\beta_j\approx\beta_{\text{max}}\sin\left(\omega_{\text{ac}} t_j+\theta \right)$, where $\beta_{\text{max}} \approx \gamma_n B/\Omega \ll 1$. 

Now, we study the effect of $\beta_j$ on the widely used XY-8 sequence~\cite{Souza12} by tuning the ratio $\alpha = \omega_{\text{DD}}/\omega_{\text{ac}}$. The ideal signal after a single application of the XY-8 sequence is given in Appendix~\ref{app:idealSig}, Eq.~\eqref{eq:idealSignal}. This ideal signal is completely independent of the initial phase $\phi$ of the sensor spin. In fact, for $\beta_\text{max}\neq0$ the influence of the tilt arises in higher orders of $\beta_\text{max}$, which we characterize in the following.
As it can be seen in Fig.~\ref{fig:harmonicaxis}(b) we apply the DD sequence such that the tilt of the axis is maximal, which we expect to be the worst possible case, therefore we set  $\theta=0$. 
In addition we have that for $\alpha=1$ and equally-spaced pulses, a constant magnitude for $\beta_j$  up to a sign change, i.e., $|\beta_j| = |\beta_{\text{max}}|$ $\forall j$, see Fig. \ref{fig:harmonicaxis}(b). Hence, during  pulses the state is rotated around the axis 
\begin{eqnarray}
\hat{n}_j^{\alpha=1} = \left(\cos\varphi_j\cos\beta_{\text{max}},\sin\varphi_j\cos\beta_{\text{max}},(-1)^j\sin\beta_{\text{max}}\right)^T.
\end{eqnarray}
We apply an XY-8 sequence with 8 pulses and find that for small tilting angles $\beta_{\text{max}}$ we have
\begin{eqnarray}\label{alpha1}
P_{\alpha=1} = - 16 \left[ \sin \left( \frac{2\gamma_n B}{\omega_{\text{ac}}}-2\phi \right) -1 \right] \beta_{\text{max}}^6 + O(\beta_{\text{max}}^7).
\end{eqnarray}
The result in Eq.~(\ref{alpha1}) tells us that when the frequency $\omega_{\text{DD}}$ is tuned to  $\omega_{\rm ac}$ (note that $\alpha = 1$)  the sensor is only  marginally affected by the presence of the tilting angle. Therefore, a change on the initial phase $\phi$ would have almost no effect on the observed spectrum. 

In contrast, for a decoupling frequency such that $\omega_{\text{DD}}=2\omega_{\text{ac}}$ we have that the rotation axis is 
\begin{eqnarray}\label{axis2}
\hat{n}_j^{\alpha=2}=\begin{pmatrix}
\cos\varphi_j\cos\beta_j \\ \sin\varphi_j\cos\beta_j \\ (-1)^{2\mathrm{mod}\lfloor (j-1)/2 \rfloor}\sin\beta_j
\end{pmatrix}.
\end{eqnarray}
and after the application of an XY-8 sequence we obtain 
\begin{eqnarray}\label{alpha2}
	P_{\alpha=2} &=& 8 \left\{\cos \left(\frac{\gamma_n B}{\sqrt{2}\omega_{\text{ac}}}\right)^2 \times\right. \notag \\ 
	&&\left. \left[1+\sin\left( \frac{2(\sqrt{2}-1)\gamma_n B}{\omega_{\text{ac}}}+2\phi\right)\right]\right\}\beta_{\text{max}}^2+O(\beta_{\text{max}}^3).\nonumber\\
\end{eqnarray}
Here  the signal is already affected by the square of the tilting angle which is the reason for a spurious resonance to appear. 

For the fourfold frequency, $\omega_{\text{DD}} = 4 \omega_{\text{ac}}$, and $\gamma_nB/\omega_{\text{ac}}\ll1$ we have that the transition probability is 
\begin{eqnarray}\label{alpha4}
	P_{\alpha=4} \approx 2\left(\sqrt{2}-2\right)\left[\sin(2\phi)-1\right]\beta_{\text{max}}^2+O(\beta_{\text{max}}^3),
\end{eqnarray}
which also contains the second order of $\beta_{\text{max}}$ and as it can be seen in Fig.~\ref{fig:harmonicaxis} (d) the rotation axis corresponding to consecutive $\rm X$ (or $\rm Y$) pulses are always different.

In Fig.~\ref{impactsecorder} (a)  we have analytically computed (see Eq.~\eqref{eq:secOrder} in the Appendix~\ref{app:secOrder}) the impact of the phase $\phi$ of the initial spin state on the factors accompanying the second order on the tilting angle for different values of the $\omega_{\rm DD}$ frequencies after the application of a single XY-8 sequence under the assumption $\gamma_nB/\omega_{\text{ac}}\ll1$. From that figure we can extract important conclusions. On the one hand for the frequencies $\omega_{\rm DD}$ with  $k=1$ and $1/\alpha=l = 1, 3, 5, \ldots$, there is no  dependence on $\beta_{\text{max}}^2$. Hence these resonances are independent up to the order $\beta_\text{max}^6$, therefore the effect of $\phi$ is entirely negligible for short pulses. Note that for the case $\theta=\pi/2$ we will find no spurious contribution as the pulses are located on the nodes of the ac-field thus we have $\beta_\text{max}=0$. In addition, for the cases $l=3,5,\dots$ the collected error is completely equivalent to the $l=1$ case as the field in the moment of pulse application is exactly the same. On the other hand, for other values of  $\alpha$  with  spurious resonances, the dependence on $\phi$ can be clearly observed. Furthermore the vertical lines in Fig.~\ref{impactsecorder} (a) correspond to the cases $\alpha = 1, 2, 4$ that we have previously discussed in Eqs.~(\ref{alpha1}), (\ref{alpha2}) and (\ref{alpha4}) respectively, and a numerical  check shown in Fig. \ref{impactsecorder} (b) stresses the agreement with those analytical expressions. In addition, Fig.~\ref{impactsecorderthreeapplications} in the Appendix shows the equivalent for Fig.~\ref{impactsecorder} (a), but for three applications of the XY-8 sequence. Here, the phase dependent accumulation is even more pronounced while the width of the resonances is decreased.

This dependence on the phase $\phi$ motivates the development of a criterion to identify spurious resonances.
\begin{figure}[t!]
\begin{centering}
\vspace{0cm}
\hspace{-0.3cm} \includegraphics[width=0.95\columnwidth]{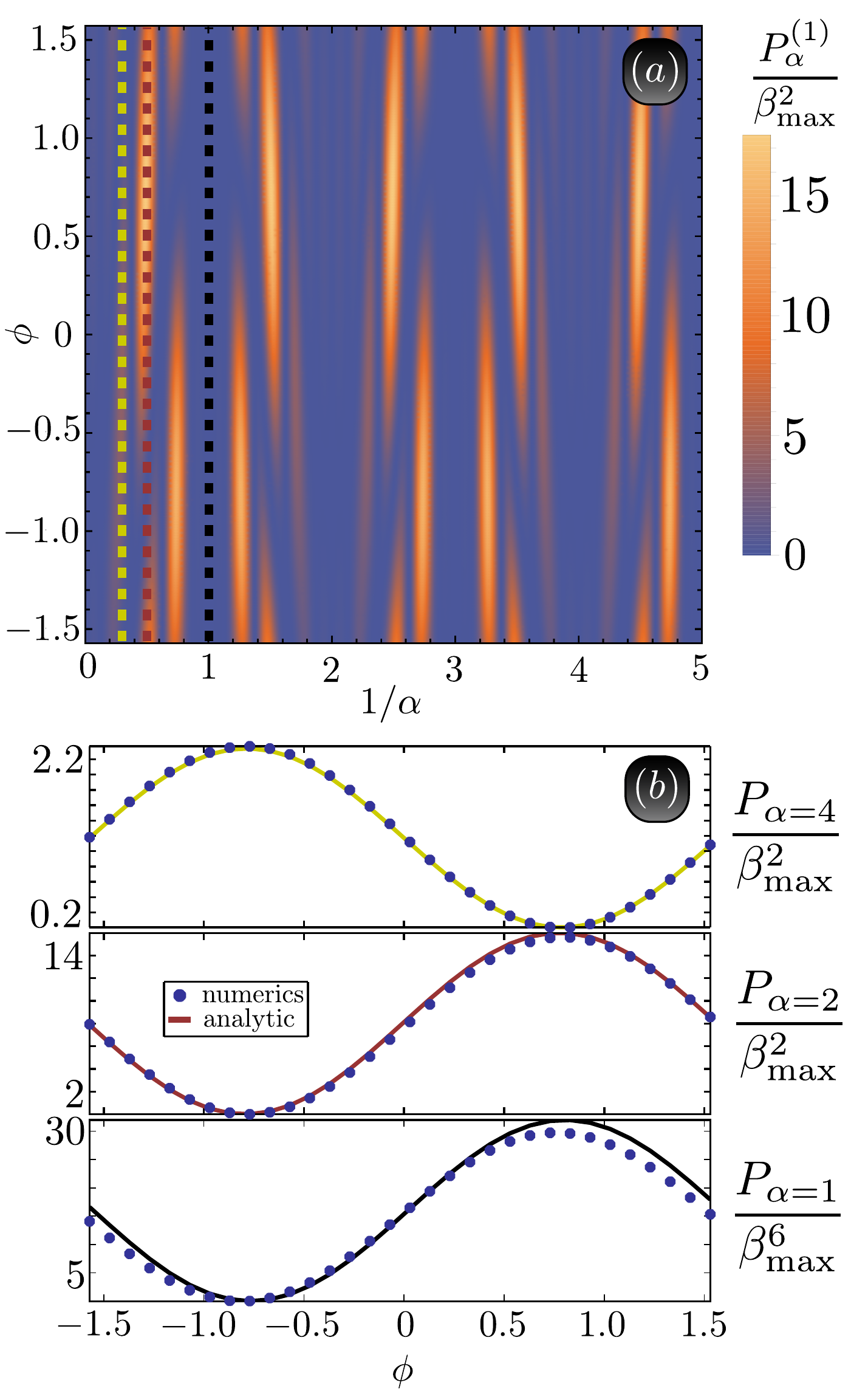}
\par\end{centering}
\caption{(a) Impact of the second order in the tilting angle for the XY-8 sequence as a function of $1/\alpha$ and the phase angle $\phi$. The cuts indicated by the dashed lines have different values of $\alpha$ and are shown in (b), where we compare the analytic results (solid lines) of Eqs.~(\ref{alpha1}), (\ref{alpha2}), and (\ref{alpha4}) with a numerical simulation (dots) of the behavior of the transition probability $P$ under the Hamiltonian given in Eq.~(\ref{eq:Hamiltonian}) for $\omega_{\text{ac}}=2\pi\times1\,\mathrm{MHz}$, $\gamma_n B=2\pi\times0.12\,\mathrm{MHz}\,$, $\theta=0$, and $\beta_{\text{max}} = 0.012$.} \label{impactsecorder}
\end{figure}

According to the behavior predicted by Eqs.~(\ref{alpha1}), (\ref{alpha2}) and (\ref{alpha4}) we are able to suppress and 
enhance the quadratic order in $\beta_{\text{max}}$ by choosing a suitable value of the initial phase $\phi$ when the resonances are spurious. Therefore, we can detect a spurious resonance by the oscillation of its associate peak's height when choosing different initial phases of the state. More specifically, after the a first experiment and the recording of the spectrum $P_{\phi_1}$, we repeat the experiment with a different initial phase to obtain $P_{\phi_2}$. In this manner for every real resonance we will have
\begin{eqnarray}
P_{\phi_1} = P_{\phi_2} + O{(\beta_{\text{max}}^6)} \approx P_{\phi_2} , 
\end{eqnarray}
meaning that the effect of the tilting angle is negligible. On the contrary, spurious resonances differ already at the order  $\beta_{\text{max}}^2$. Hence by comparing  $P_{\phi_1}$ and $P_{\phi_2}$ the real resonances can be identified.
A further improvement concerning  resolution on spurious peaks can be made by recording multiple initial phases to construct a spectrum of spurious resonances. In this respect one can define the following quantity 
\begin{eqnarray}
W = \max_{\phi_i,\phi_j} \left|P_{\phi_i}-P_{\phi_j}\right|,
\label{eq:witness}
\end{eqnarray} 
where the maximum is taken over all recorded initial phases. $W$ contains all the peaks but the real resonances because in this case $P_{\phi_i} \approx P_{\phi_j}$ $\forall \phi_i, \phi_j$ leading to $W\approx 0$. 
It is important to stress that this criterion is one-directional namely in the case of having multiple ac-fields with frequencies $\omega_{ac, j}$ and some of them are integer multiples of another one, i.e. $\omega_{ac,k} = \mu \omega_{ac,l}$, the real resonances $\omega_{ac,l}$  can not be distinguished from the spurious contribution of $\omega_{ac,k}$.

\subsection{Effects of pulse errors.}
We derived the above criterion for pulses which are only disturbed by the action of the ac-field during the pulse time. However, a real situation will also suffer from the presence of a detuning $\Delta$, see Eq.~(\ref{classical}), and flip-angle errors caused by fluctuations in the Rabi frequency $\Omega$ in $H_c$. For the following analysis we will consider static errors in $\Delta$ and $\Omega$. Note that  this condition can be justified by assuming that both $\Delta$ and $\Omega$ are slowly varying. The detuning $\Delta$ of the applied control field from the transition frequency of the sensing qubit, see Eq.\eqref{eq:Hamiltonian},  tilts the rotation axis out of the $x-y$ plane by an angle $\gamma$ that can be quantified as $\gamma=\tan^{-1}(\Delta/\Omega) \approx \Delta/\Omega$ if  $\Delta \ll \Omega$. In addition an error on $\Omega$ results in non perfect $\pi$ pulses with the angle of rotation $\pi+\delta$. To analyze the signal in small control errors, we write all possible deviations as $\beta_{\text{max}}=\tilde{\beta}\eta$, $\gamma=\tilde{\gamma}\eta$, $\delta=\tilde{\delta}\eta$ and expand the signal with respect to the small parameter $\eta$. In the case of ideal control, $\eta\rightarrow 0$. Note that different errors are described by the independent proportionality constants, $\tilde{\beta}$, $\tilde{\gamma}$, and $\tilde{\delta}$. A repetition of the calculation for finding Eq.~(\ref{alpha1}) yields

\begin{eqnarray}
	P_{\alpha=1} &=& \frac{1}{4}\left[4\tilde{\beta}^2-4\tilde{\gamma}^2+\tilde{\delta}^2\right]^2 \times \notag \\
	&&\left[\left(2\tilde{\beta}+\tilde{\delta}\right)\cos \phi + \left(2\tilde{\beta}-\tilde{\delta}\right)\sin\phi)\right]^2\eta^6 \notag \\
	&&+O(\eta^7),
\end{eqnarray}
while it can be shown that $P_{\alpha=2}$ and $P_{\alpha=4}$ do not change in the second order of $\eta$. Therefore, our criterion is valid to identify spurious peaks under the presence of error sources.  

\subsection{Remark on pulse phases}
It is worth to mention that the preparation of the initial state in the $x$-$y$ plane with different initial phases $\phi_j$, or different choices of the rotation axis for the decoupling pulses are interchangeable. The later can be achieved by a variation of $\vartheta$ in Eq.~\eqref{eq:ControlHamiltonian}. More specifically, a preparation in $\rho=|+_x\rangle \langle+_x|$ and choosing the rotation axis $X_{\vartheta_j}$ and $Y_{\vartheta_j} = X_{\vartheta_j+\frac{\pi}{2}}$ is equivalent to the situation described throughout the paper. When both phases are changed, the equations above still hold if one makes the identification $\phi\mapsto\phi-\vartheta$. 

\section{Detection of a quantum signal}\label{quantum}
\subsection{A scheme for  quantum emitters}
In a quantum setting  the classical field is replaced by one or more nuclei, each of them oscillating at its own Larmor frequency, and coupled  differently to the sensor spin. For the sake of simplicity, we stick to a single remote spin. In that case the free evolution of the system is dictated by the Hamiltonian
\begin{eqnarray}
	H = \omega_c\sigma_z  - \omega I_z + \vec{\sigma} \tilde{A} \vec{I} \approx \omega_c\sigma_z  - \omega I_z + \sigma_z \vec{A} \vec{I},
\end{eqnarray}
where $\tilde{A}$ is the hyperfine tensor describing the interaction between the sensor and the target spin. We assumed that the energy splitting is much larger than the interaction with each remote spin, $\omega_c \gg \omega$, therefore the central spin does not flip and we applied the secular approximation which removes the corresponding flip-flop terms. In a rotating frame of the free energy terms $ \omega_c\sigma_z  - \omega I_z$ we obtain the following Hamiltonian
\begin{eqnarray}
	H_I(t) = \sigma_z  \left[A_x \cos(\omega t) I_x - A_y \sin(\omega t) I_y + A_z I_z \right]. \label{eq:quantumHamiltonian}
\end{eqnarray}
Hence the levels of the central spin are shifted by the amplitudes $A_i$ which are the analogue to the amplitude of the classical field. Note that the first two contributions at the right hand side of Eq.~\eqref{eq:quantumHamiltonian} reassemble the cases $\theta=0$ and $\theta=\pi/2$ simultaneously. Thus, if we want to use our criterion for the identification of spurious resonances  we have to ensure that $|\vec{A}|\ll\Omega$ which is the condition giving rise to small tilting angles. 

The regime where this  condition holds is easily satisfied in NV based schemes, that we will comment below, as typical couplings to remote spins are around $2\pi\times20\,\mathrm{kHz}\frac{\mathrm{nm}^3}{r^3}$ (with $r$ being the distance between the NV center and each nuclear spin), while  driving frequencies can be easily selected around $2\pi\times 30\,\mathrm{MHz}$.

\subsection{Numerical results in NV-based schemes}\label{results}
\begin{figure*}[t!]
\begin{centering}
\vspace{0cm}
\hspace{-0.3cm} \includegraphics[width=1.8\columnwidth]{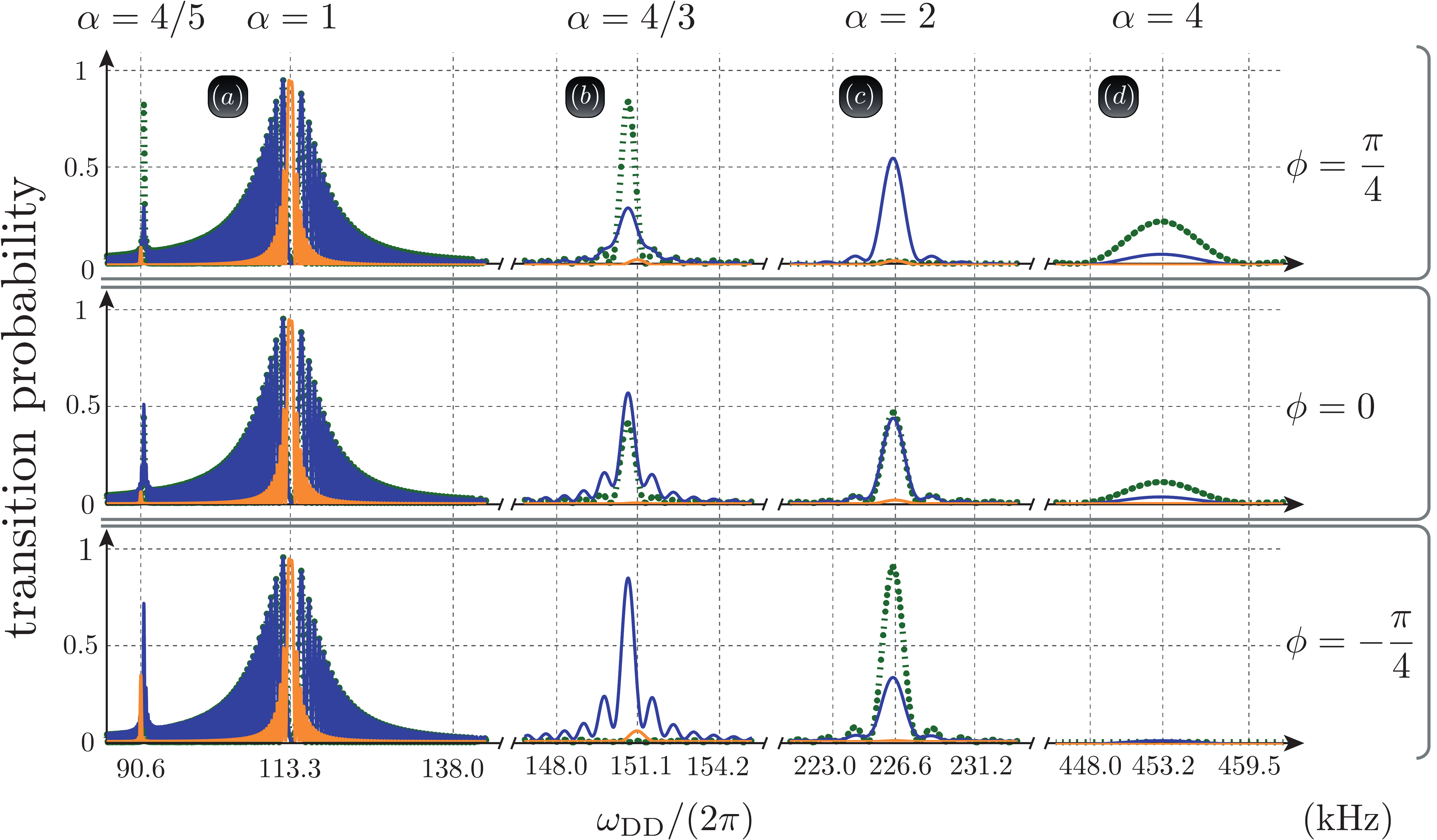}
\par\end{centering}
\caption{Simulation of $P_\phi$ for a NV center coupled to a single spin employing the XY-8 [blue (solid dark grey)] and AXY-8 [orange (solid light grey)] sequences and different initial phases of $\pi/4,\,0,\,-\pi/4$ in each row from top to bottom. The graphs in column (a) show the large resonance for $\alpha=1$ and the smaller $\alpha=4/5$ resonance. In the columns (b), (c) and (d) the spurious peaks for $\alpha = 4/3,\, 2,\, 4$ are displayed respectively. The $\alpha=1$ resonance corresponds to a sensing time of $T\approx2.4\,\mathrm{ms}$. The green (dark grey) dashed curves represent an XY8 sequence but with no error in the Rabi-frequency.} \label{PhaseCompare}
\end{figure*}
A widely used sensor spin  corresponds to an NV center in diamond which qualifies itself through through long decay and coherence times even at room temperatures~\cite{Doherty13, Dobrovitski13}. 
The Hamiltonian of an NV center and its surrounding nuclear spins without control reads
\begin{eqnarray}\label{NVH}
	H = DS_z^2 - \gamma_e B_z S_z -\sum_j \gamma_j B_z I_j^z + S_z\sum_j \vec{A}_j\cdot \vec{I}_j,
\end{eqnarray}
where $B_z$ represents an external magnetic field applied along the NV axis, the $\hat{z}$-direction, $D=2\pi\times2.87\,\mathrm{GHz}$ is the zero field splitting, $\gamma_e$,$\gamma_j$ are the electronic and nuclear gyromagnetic ratios respectively, and $\vec{A}_j$ is the hyperfine vector describing the dipolar interaction between the spin-1 NV center and the $j$-th remote spin-$1/2$ nuclei ($S$ and $I$ represent the spin-1 and spin-1/2 operators respectively). The Hamiltonian in  Eq.~(\ref{NVH}) has been cast in the secular approximation where all terms allowing flip-flop dynamics of the NV center's electron spin have been removed. Note that this approximation is well justified because of the large values of energy mismatch. 
We restrict to the  subspace containing only the electronic spin states $|m_s=0\rangle,\, |m_s=1\rangle$ which we choose as our sensing qubit~\cite{Doherty13, Dobrovitski13}. By using $|1\rangle\langle1|=(\sigma_z+1)/2$ and by going to the rotating frame of NV electron spin we arrive at the Hamiltonian under control
\begin{eqnarray}
H^\prime = \sum_j \vec{\omega}_j\cdot\vec{I}_j+\frac{\sigma_z}{2}\sum_j\vec{A}_j\cdot\vec{I}_j- \Delta\frac{\sigma_z}{2} +H_c. 
\label{eq:effectiveHamiltonian}
\end{eqnarray}
where every nuclear spin rotates with its own larmor frequency $|\vec{\omega}_j|=\left|\frac{1}{2}\vec{A}_j-\gamma_jB\hat{z}\right|$. 
The control Hamiltonian $H_c$ under rotating wave approximation is described by Eq. (\ref{eq:ControlHamiltonian}).

In Fig.~\ref{PhaseCompare} we illustrate the oscillation of spurious peaks which we use for their detection and discrimination from real peaks. We present the spectrum that results from the interaction of an NV center  with a remote $^{13}$C spin ($\gamma_{C}=2\pi\times1.0705\,\mathrm{kHz}/\mathrm{G}$) at a distance of $r\approx1.19\,\mathrm{nm}$ from the NV center and located in one of the available diamond lattice positions. This gives rise to a hyperfine coupling $\vec{A}=2\pi\times(15.0\;6.4\;11.9)^T\,\mathrm{kHz}$. The applied field strength of the external magnetic field reads $B_z=100\,\mathrm{G}$. Concerning the possible error sources  we have taken into account that the nitrogen atom inherent to the NV center might change the energy splitting of the electronic spin due to a hyperfine interaction of up to $\sim 2\pi \times1\,$MHz~\cite{Loretz15,Doherty13} when the intrinsic nitrogen spin is not polarized. The detuning is stable because of the long $T_1$ time of the nitrogen spin. Therefore, we choose $\Delta = 2\pi\times1\,\mathrm{MHz}$ in our numerical simulations and include a relatively large $3\%$ error in the Rabi-frequency $\Omega$ which  is set to be $2\pi\times30\,\mathrm{MHz}$. We compare the spectra obtained for AXY-8 which is a robust sequence suitable for quantum computing and sensing \cite{Casanova15,Wang15,Casanova2016,Wang2016} (see Appendix~\ref{app:AXY}) and XY-8 sequences for $N=70$ repetitions of the corresponding protocols, meaning 2800 pulses for AXY-8 and  560 pulses for XY-8. The AXY sequences provide improved sensing resolution in a way similar to the proposal in Ref. \cite{Zhao2014} that has been experimentally verified in \cite{Ma2015}. In addition, the AXY sequence utilize the robust composite Knill pulses \cite{Ahmed13,Ryan2010} to compensate pulse errors, which is important when the number of applied pulses is large.

We choose $\Omega(t)$ in a way such that the AXY-8 sequence is assembled with  $f_1=4/(5\pi)$ (see Appendix), while the coefficient for XY-8 is always fixed, for the first harmonic contribution, to $f_1=4/\pi$. We run the simulation with three initial phases $\phi=0$ and $\phi=\pm\pi/4$. The important parts of the spectra are shown in Fig.~\ref{PhaseCompare}. We can clearly distinguish the spurious peaks from the real peaks. The spurious resonances in Fig.~\ref{PhaseCompare} (b), (c) and (d) change the peak heights under the varying initial phase which makes them easy to detect. The AXY-8 sequence shows less amplitude in the $\alpha >1$ spurious peaks, as it reduces the effective coupling to the remote spins by $f_1$ (a fraction of $1/5$ as the one of XY-8) and therefore reduces the tilting angle, in addition it employs rotations around 6 axis instead of 2 as XY-8 does and is therefore more robust against the accumulation of the fake signal. Also see the green (dark grey) dashed line which illustrates the sensitivity of XY8 with respect to errors in the Rabi-frequency in comparison with the blue (dark grey) line. The occurrence for the $\alpha=4/3$ resonance (Fig.~\ref{PhaseCompare} (b)) in the AXY-8 sequence is due to the large detuning and the high peak at $\alpha=4/5$ Fig.~\ref{PhaseCompare} (a) results because of the larger fourier coefficient for $f_5$ when compared to a standard XY sequence with equally spaced pulses. However, this resonance is also easy to detect.
 
\subsection{Distinguishing close peaks.}
\begin{figure}[t!]
\begin{centering}
\vspace{0cm}
\hspace{-0.3cm} \includegraphics[width=0.95\columnwidth]{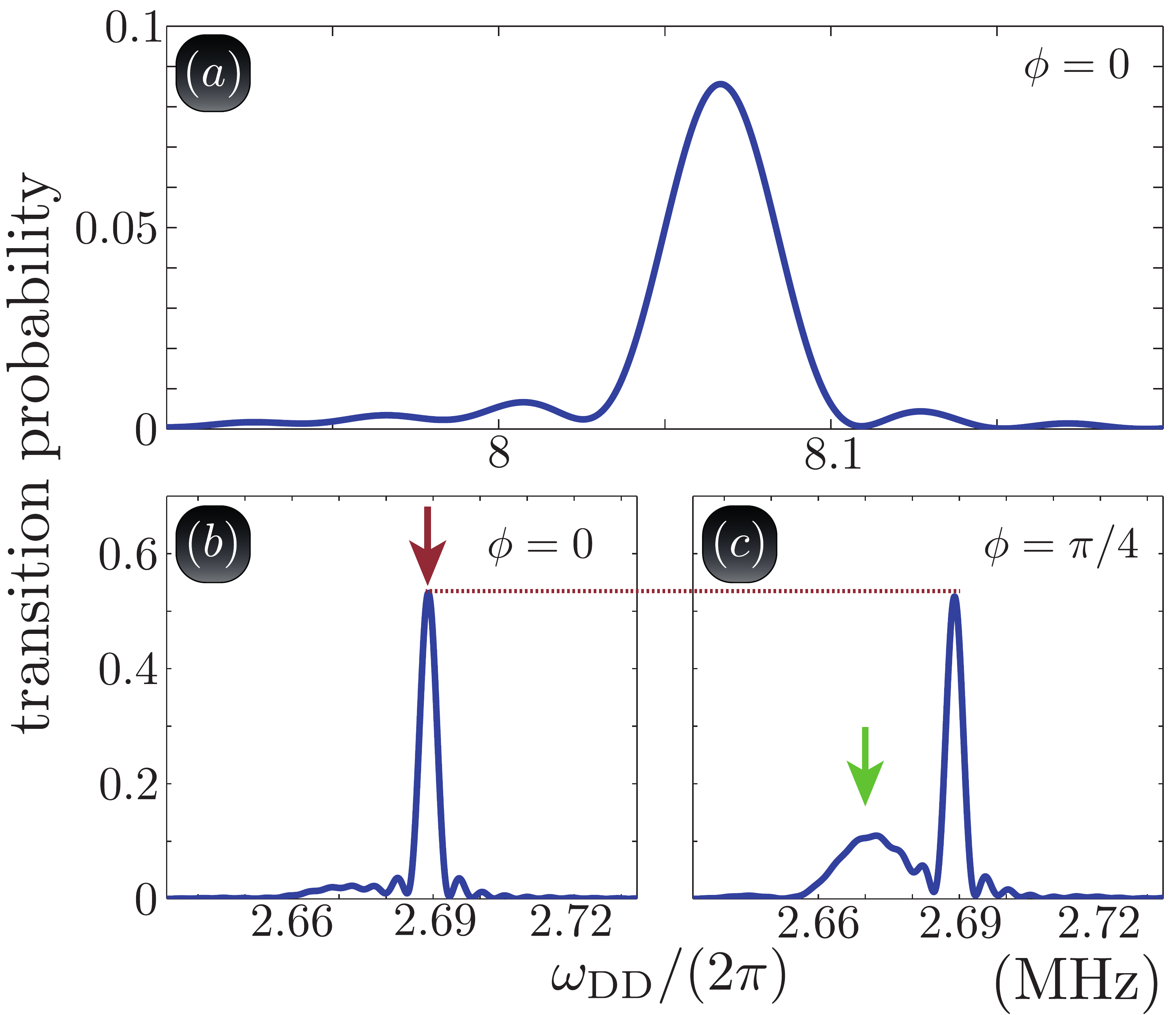}
\par\end{centering}

\caption{Simulation of $P_\phi$ for an NV center with a hydrogen and a strongly coupled carbon spin under the control of AXY sequences. a) and b) show results for $f_1=4/(1.2\pi)$ and $f_3=4/(1.2 \pi)$ respectively. The arrow in (b) indicates the position of the hydrogen resonance, while the arrow in (c) marks the $^{13}$C resonance. (b) and (c) show the effect when changing the initial phase from $0$ to $\pi/4$. All calculations are performed with $\Delta=2\pi\times1\,\mathrm{MHz}$, $\Omega=2\pi\times20\,\mathrm{MHz}$ and $\delta=0.03\times\Omega t_{\text{flip}}$. Here, 1920 decoupling pulses are used which corresponds to a sensing time $T\approx23.9\,\mathrm{\mu s}$ for the $\alpha=1$ and $T\approx71.7\,\mathrm{\mu s}$ for the $\alpha=1/3$ resonance of the hydrogen atom.}
\label{carbonhydrogen}
\end{figure}
Spurious resonances can induce false identification of detected nuclear spins~\cite{Loretz15}. For example, ${^1}$H has a gyromagnetic ratio of $\gamma_{^1{\rm H}} = 2\pi\times4.2576 \,\mathrm{kHz}/\mathrm{G}$ thus with $B_z=600\,\mathrm{G}$ we expect a resonance peak at $\approx 2\pi\times2.555\,\mathrm{MHz}$. Unfortunately, NV center based detection of hydrogen suffers from the natural occurrence of $^{13}$C spins in the diamond lattice~\cite{Loretz15}. These carbon spins produce a spurious resonance peak at approximately $1.0057$ times the resonance frequency of the hydrogen spin since their Larmor frequency at this field is around $2\pi\times0.642\,\mathrm{MHz}$. Hence their $\alpha=4$ resonance will appear around the Larmor frequency of the hydrogen. As long as the absence of $^{13}$C is not ascertained by independent means, the detection of hydrogen can not be achieved unambiguously. We use our recently introduced AXY-8 sequence and the above defined criterion to identify the spurious resonances. 

For the following simulations we are guided by the data presented in \cite{Loretz15}. Here, a hydrogen spin with $\vec{A}_\mathrm{H}=2\pi\times(14.5\;0\;500)^T\,\mathrm{kHz}$ is considered. We assume a carbon spin with $\vec{A}_{C} = 2\pi\times(103,\;103,\;73)^T\,\mathrm{kHz}$ $(r_{C}\approx0.59\,\mathrm{nm})$ at one of the possible positions in the diamond lattice surrounding the NV center. The magnetic field is tuned to $B_z=1836\,\mathrm{G}$. 

Fig.~\ref{carbonhydrogen} (a) shows the transition probability in the region of the expected hydrogen resonance for $\alpha=1$, ($k=l=1$) and $f_1=4/(1.2\pi)$ using the parameters mentioned below the plot. From here it is not clear, to which element this peak has to be assigned, whether this peak indeed represents a real resonance or if it spurious. Increasing the selectivity of the AXY-8 sequence by changing the sequence to $\alpha=1/3$ ($k=1,\,l=3$) and changing to $f_1=0,\,f_3=4/(1.2\pi)$ leads to the spectrum (b) which clearly shows the resonance peak of the hydrogen and marks the spurious $^{13}$C resonance, which can be identified undoubtedly by changing the phase as shown in Fig.\ref{carbonhydrogen} (c). Again, note that even if there would be no  $^{13}$C present, the constant height of the hydrogen peak under the phase cycling proofs that the peak is a real resonance of a present interacting spin.

\section{Conclusion}
We have defined a criterion that allows the identification of spurious resonances as they appear in widely used dynamical decoupling schemes of the XY-family which can easy be implemented in existing experimental setups as it only requires a phase change of the applied pulses. To understand its working mechanism, we calculated the effect of a XY-8 decoupling sequence for detection of a single classical ac-field and motivated the definition by the different leading orders of the tilting angle of the rotation axis, which is responsible for the appearance of spurious resonances. A further calculation verified the validity of the criterion in a quantum setting under the sufficient condition of a strong enough driving field used for the $\pi$-pulses. Later, we applied the criterion to NV center coupled to a single spin where we illustrated the working principle. For a second example, we solved the detection uncertainty of hydrogen atoms when using NV centers by employing the AXY-8 sequence. 

\section{Acknowledgments}
This work was supported by an Alexander von Humboldt Professorship, the ERC Synergy grant BioQ and the EU Projects EQUAM and DIADEMS. J.C. acknowledges support to the Alexander von Humboldt foundation.

\begin{appendix}
\renewcommand\thefigure{\thesection.\arabic{figure}}
\setcounter{figure}{0}
\section{Ideal signal after a single application}
\label{app:idealSig}
The definitions of the hamiltonians and sequences in sec.~\ref{motivation} allow us to calculate the first order of the signal in $\beta_\text{max}$. It turns out, that the corresponding result is independent of the tilting angle to first order, hence represents the ideal signal after the application of a single unit of the XY-8 sequence (8 pulses):
\begin{eqnarray}
P_{\alpha,\theta}^{(1)} &=& \frac{1}{2} \Bigg \lbrace 1- \cos \Bigg [ 16 B\gamma_n \Bigg ( \cos\left( \frac{3\pi}{4\alpha}\right)+ \cos\left( \frac{5\pi}{4\alpha}\right) \notag \\ 
&&+ \cos\left( \frac{11\pi}{4\alpha}\right) + \cos\left( \frac{13\pi}{4\alpha}\right) \Bigg ) \notag \\
&&\times \frac{ \sin\left( \frac{\pi}{4\alpha}\right)^3\sin\left(\frac{4\pi}{\alpha}+\theta\right)}{\omega_\text{ac}} \Bigg] \Bigg \rbrace + O(\beta_\text{max}^2).
\label{eq:idealSignal}
\end{eqnarray}
Note that the zero-th order contribution in $\beta_\text{max}$ (i.e., $P_{\alpha,\theta}^{(1)}$ with $\beta_\text{max}=0$) is independent of $\phi$. However, higher orders on $\beta_\text{max}$ can provide a dependence on $\phi$. For $\beta_\text{max}=0$, this equation represents the ideal signal. 

\section{Impact of the second order}
\label{app:secOrder}
\begin{figure}[htp!]
\begin{centering}
\vspace{0cm}
\hspace{-0.3cm} \includegraphics[width=0.95\columnwidth]{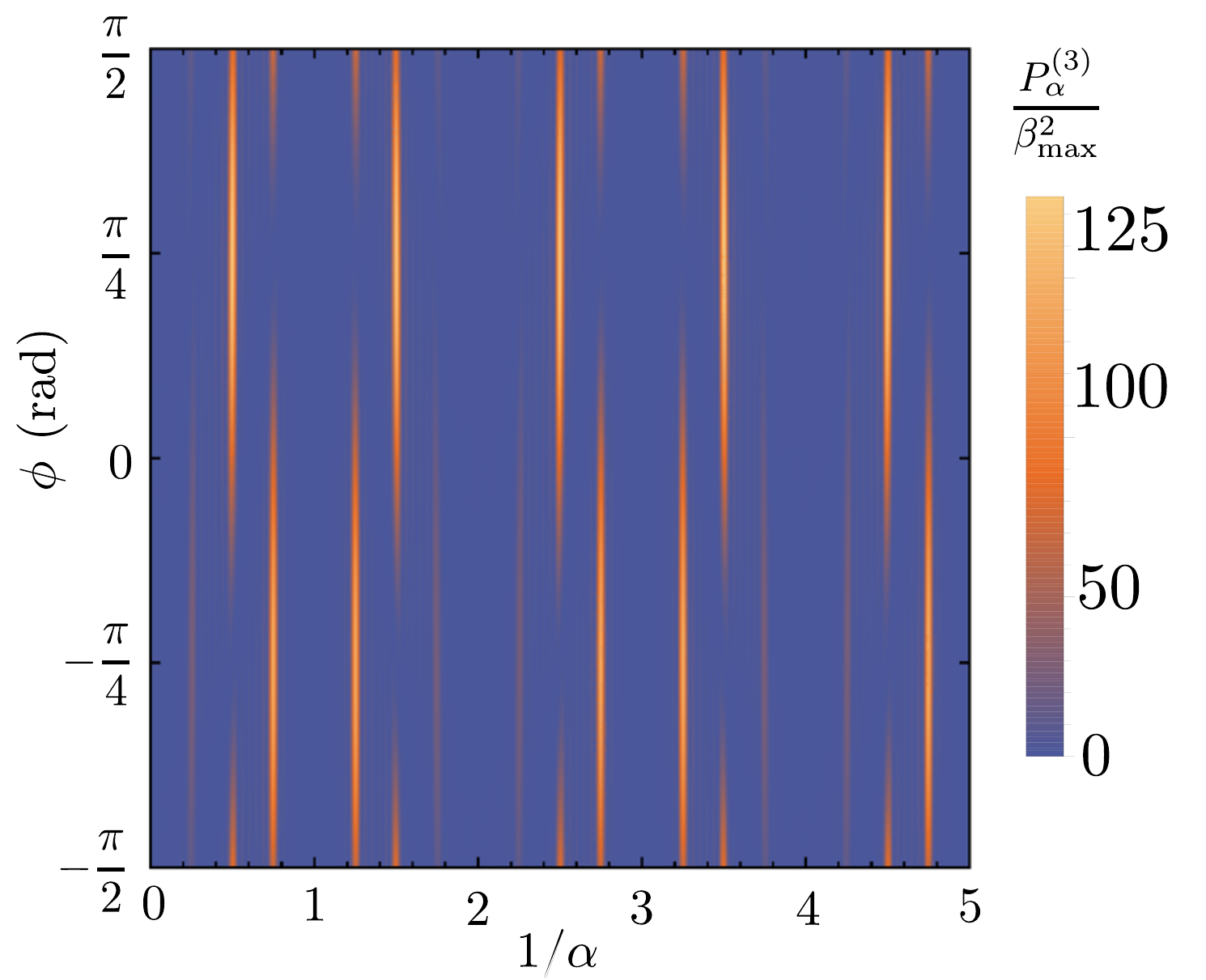}
\par\end{centering}
\caption{Impact of the second order after three applications of the XY-8 sequence. The values for the figure are calculated using Eq.~\eqref{eq:thirdOrder}} \label{impactsecorderthreeapplications}
\end{figure}
With the definitions given in sec.~\ref{motivation}, we can calculate the spectrum for a single application of the XY-8 sequence (8 $\pi$-pulses) as
\begin{eqnarray}
P_{\alpha}^{(1)}&\approx&\Bigg \lbrace  \Bigg[ \sin\left(\frac{\pi}{2\alpha}\right) -\sin\left(\frac{5\pi}{2\alpha}\right) +\sin\left(\frac{11\pi}{2\alpha}\right) \notag \\ &&-\sin\left(\frac{15\pi}{2\alpha}\right)\Bigg] \cos\phi \notag \\ 
&&+ \Bigg[\sin\left(\frac{3\pi}{2\alpha}\right) -\sin\left(\frac{7\pi}{2\alpha}\right) +\sin\left(\frac{9\pi}{2\alpha}\right) \notag \\ &&-\sin\left(\frac{13\pi}{2\alpha}\right) \Bigg] \sin\phi \Bigg \rbrace ^2 \beta_{\text{max}}^2 + O(\beta_{\text{max}}^3),
\label{eq:secOrder}
\end{eqnarray}
which is valid under the assumption $\gamma_nB/\omega_{\text{ac}}\ll1$. This result gives $P_{\alpha=3}^{(1)}/\beta_{\text{max}}^2\approx9\cos{\phi}^2/4$ and will thus oscillate under a changing initial phase. However, after three applications of the XY-8 sequence (24 $\pi$-pulses), we obtain the signal
\begin{eqnarray}
P_{\alpha}^{(3)}&\approx&\Bigg \lbrace 16 \cos\left(\frac{\pi}{4\alpha}\right)^4 \Bigg[ \sin\left(\frac{7\pi}{2\alpha}\right)-\sin\left(\frac{9\pi}{2\alpha}\right)+\sin\left(\frac{23\pi}{2\alpha}\right)\notag \\&&-\sin\left(\frac{25\pi}{2\alpha}\right)+\sin\left(\frac{39\pi}{2\alpha}\right)-\sin\left(\frac{41\pi}{2\alpha}\right)\Bigg]^2 \notag \\ 
&&\times \Bigg[ \cos\phi + 2\cos\left(\frac{2\pi}{\alpha}\right)\cos\phi - \sin \phi \notag \\&&+ 2\cos \left(\frac{\pi}{\alpha}\right) \left(\sin\phi-\cos\phi\right) \Bigg]^2 \Bigg \rbrace \beta_{\text{max}}^2 + O(\beta_{\text{max}}^3).
\label{eq:thirdOrder}
\end{eqnarray}
Interestingly, for $\alpha=3$ we have $P_{\alpha}^{(3)}\approx O(\beta_{\text{max}}^3)$ . The same calculation can be done for other odd numbers of $\alpha>1$. Consecutive applications show that after $n\alpha$ sequences, for $\alpha>1$ and $n\in \mathbb{N}$, the corresponding signal is again zero, thus for these resonances no accumulation of phase is accomplished. 
Fig.~\ref{impactsecorderthreeapplications} shows Eq.~\eqref{eq:thirdOrder} for different values of $1/\alpha$ and $\phi$. This sequence requires three times the evolution time as used for Fig.~\ref{impactsecorder}, thus the peaks are much narrower and it can be observed how the spurious signal accumulation is only present at certain relations of $\omega_\text{DD}/\omega_\text{ac}$ while it is highly phase dependent.
\section{AXY pulse sequence}
\label{app:AXY}
The AXY-8 pulse sequence as presented in \cite{Casanova15} is an extension to the XY-familiy. Here, each X(Y)-pulse is replaced by five pulses which form a composite X(Y)-pulse. The five sub-pulses are non equally spaced but the spacing is symmetric around the 3rd pulse and they obey a specific phase relation similar to the Knill-sequence \cite{Ryan2010,Ahmed13}, making the sequence highly robust against pulse errors. In addition, the sequence allows for single spin addressing as under resonance $\omega=2\pi l f_{\text{DD}}$ with $\omega$ the larmor frequency of the target spin, it dictates an evolution with the effective Hamiltonian
\begin{eqnarray}
H = \frac{m_s}{4}f_{l}\sigma_z\vec{I}\cdot\vec{a} ,
\end{eqnarray}
where $m_s=\pm1$ the spin quantum number of the NV center electron spin-1, $\vec{a}$ the effective coupling vector and $f_{l}$ the Fourier coefficient of the $l^{\mathrm{th}}$ term in the Fourier representation of the applied filter function. By changing the interpulse spacing of the introduced composite pulses, the first four coefficients can be controlled as $f_1=\xi,\,f_2=0,\,f_3=0,\,f_4=0$ where $\xi\pi\in (-8\cos\frac{\pi}{9}+4,8\cos\frac{\pi}{9}-4)$ and the corresponding pulse times $x_i\frac{T}{2}$ are given by
\begin{eqnarray}
x_{1,2}&=&\frac{1}{2\pi}\arctan\frac{\pm(2\xi\pi-12)w_1+\sqrt{3w_2}}{\sqrt{6}\sqrt{w_2-96\xi w_1\pi\pm w_1^2\sqrt{3w_2}}},
\end{eqnarray}
as well as $x_3=\frac{1}{4}$ and $x_{4,5}=\frac{1}{2}-x_{2,1}$ and we defined $w_1=4-\xi\pi$ and $w_2=w_1[960-144\xi\pi-12(\xi\pi)^2+(\xi\pi)^3]$.
Another possibility is set by $f_1=0,\,f_2=0,\,f_3=\xi,\,f_4=0$ which results in the pulse times
\begin{eqnarray}
x_j&=&\frac{1}{4}-\frac{1}{2\pi}\arctan\sqrt{q_j^2-1}
\end{eqnarray}
with the same symmetry conditions as above and $q_j=4/\left[\sqrt{5+\pi\xi}+(-1)^j\right]$ and $j=1,2$ and $\xi\in\left(-\frac{4}{\pi},\frac{4}{\pi}\right)$.
These two possibilities correspond to the resonances $\alpha=1$ and $\alpha=1/3$ respectively.
\end{appendix}

\end{document}